\documentstyle[prb,aps,twocolumn]{revtex}
\begin{document}
\voffset 0.5in
\draft
\wideabs{
\title{Exchange-correlation effects in magnetic-field-induced 
superconductivity}
\author{Klaus Capelle}
\date{\today}
\address{Departamento de Qu\'{\i}mica e F\'{\i}sica Molecular, 
Instituto de Qu\'{\i}mica de S\~ao Carlos, 
Universidade de S\~ao Paulo,\\
Caixa Postal 780, S\~ao Carlos, 13560-970 SP, Brazil }
\maketitle
\begin{abstract}
Motivated by recent experiments on the organic superconductor 
$\lambda-(BETS)_2FeCl_4$ we study the Jaccarino-Peter effect (JPE), 
the earliest example of magnetic-field-induced superconductivity, 
from the point of view of current-density-functional theory.
It is found that both Meissner (diamagnetic) and Pauli (paramagnetic) pair 
breaking are suppressed by an exchange-correlation contribution to the
vector potential, arising at the sites of the magnetic ions. This explains 
a number of otherwise puzzling experimental observations, and sheds new 
light on earlier theories of the JPE.
\end{abstract}

}
\newcommand{\be}{\begin{equation}}
\newcommand{\ee}{\end{equation}}
\newcommand{\bea}{\begin{eqnarray}}
\newcommand{\eea}{\end{eqnarray}}
\newcommand{\bi}{\bibitem}
\newcommand{\ep}{\epsilon}
\newcommand{\s}{\sigma}
\newcommand{\p}{{\bf \pi}}
\newcommand{\r}{({\bf r})}
\newcommand{\rp}{({\bf r'})}

The antagonistic nature of conventional superconductivity and magnetism 
manifests itself via the pair breaking effect of external and 
internal magnetic fields.
In the case of {\it external} magnetic fields this pair breaking can be due to 
the coupling of the field to orbital currents (i.e., the Meissner effect), 
or due to its coupling to the spins (i.e., Pauli pair breaking, giving rise 
to the Chandrasekhar-Clogston limit).\cite{tinkham}
In the case of {\it internal} fields, pair breaking is due to magnetic ions 
in the lattice. These ions, too, in general make a spin and an orbital
contribution to the system's susceptibility. 
On their own, both external and internal magnetic fields are normally adverse 
to superconductivity.\cite{tinkham}

Intriguingly, under suitable circumstances these two detrimental agents,
when present simultaneously, can compensate each other and give rise 
to superconductivity at extremely high external magnetic fields.\cite{jp} 
This phenomenon, the earliest example of magnetic-field-induced 
superconductivity, is known as the Jaccarino-Peter effect (JPE). 
At the heart of the JPE is the observation that if the magnetic ions
couple ferromagnetically among each other, but antiferromagnetically 
with the conduction electrons (via the {\it exchange field} of 
traditional theories of magnetism\cite{herring}), an externally applied 
magnetic field acts in two ways on the spins of the paired particles: 
directly, by its Zeeman coupling, and indirectly, by polarizing the magnetic 
ions. Due to the antiferromagnetic exchange coupling between 
magnetic ions and conduction electrons both effects have opposite signs and 
can thus, under suitable circumstances, cancel each other. This cancellation 
is the essence of the JPE.\cite{jp,decrouxfischer}

The present paper is devoted to a re-analysis of experimental results on the 
JPE. It is pointed out that the conventional JPE scenario, just outlined,
does not completely account for the experimental observations. 
A modification of this scenario is proposed and compared with available 
experimental data.

The JPE has very recently been observed in the organic conductor 
$\lambda-(BETS)_2FeCl_4$, for magnetic fields ranging from $18$ to 
$41$ T.\cite{uji,balicas}
Before that, it had already been seen in pseudoternary chalcogenides of the
form $Sn_xEu_yMo_zS_8$, from $4-23$ T,
\cite{fischerSN1,fischerSN2,fischerSN3,fischerPB}
and there is strong experimental evidence that the JPE is also at work in 
chalcogenides of the form
$Pb_xEu_yMo_zS_8$ (at up to $70$ T),\cite{fischerPB,okudaPB}
in transition-metal systems
($Mo_{1-x}Mn_xGa_4$ at up to $7.6$ T),\cite{fischerMO,helv2}
in heavy-fermion superconductors
($CePb_3$ at up to $15$ T\cite{linHF} and $CeCu_2Si_2$ 
at about $2$ T\cite{sheikinHF1,sheikinHF2}),
and perhaps even in high-$T_c$ cuprates
($Gd_{1-x}Pr_xBa_2Cu_3O_{7-\delta}$, near $9$ T).\cite{mapleHTSC}

It is crucially important to recognize that in the standard JPE 
scenario\cite{jp,decrouxfischer} it is only the action of the external 
magnetic field on the electron {\it spins} that is compensated by the internal 
(exchange) field, not that on the {\it currents}. This imposes a severe 
restriction on the external magnetic field used for observation of the JPE: 
while it must be strong enough to cancel the exchange field arising from the 
magnetic ions, it must not, simultaneously, destroy superconductivity 
diamagnetically, by its coupling to the orbital currents. One {\it ad hoc} way
to reconcile these two constraints would be to assume that in all systems
in which the JPE has been observed the orbital upper critical field $H_{c2}$ 
is very much higher than the (already quite high) fields at which the JPE sets 
in.\cite{decrouxfischer}

In view of the recent experiments\cite{uji,balicas} on 
$\lambda-(BETS)_2FeCl_4$ it seems worthwhile to explore if this
assumption is really always necessary. In fact, in these experiments
the JPE region in the phase diagram begins at $18 $ T, while $H_{c2}$
is about $3.5$ T,\cite{balicas} so that the relation between both fields
is the opposite of what one would expect on the basis of the original
theory of the JPE. Similarly, in the
experiments\cite{fischerSN1,fischerSN2,fischerSN3,fischerPB} 
on $Sn_{1-x}Eu_xMo_6S_8$
the JPE manifests itself as a distinct phase in the temperature vs. magnetic 
field phase diagram that appears for external fields higher than $4$ T, 
whereas superconductivity first disappears, upon increasing the magnetic 
field from zero, at what appears to be a conventional orbital critical field 
of less than $1$ T.
And in $Mo_{1-x}Mn_xGa_4$ the JPE is observed at up to $7.6$ T, 
\cite{fischerMO,helv2} which is not above the orbital critical field of 
$8.6$ T but sufficiently close to it that a description in terms of spin 
effects only becomes problematic.
The question to be explored in the present paper is thus: 
How can {\it spin} compensation induce superconductivity at fields 
comparable too, or higher than, the {\it orbital} critical field?

The fact that the JPE has been observed in physically very
different systems, ranging from organic conductors and ternary
chalcogenides to heavy fermion and high-temperature superconductors, 
suggests that the answer to this question is intrinsically tied to
the physics of the JPE itself, and not a special feature of one particular
type of material.

To investigate these issues, we first restate the
compensation between the external and the internal magnetic field 
in the framework of spin-density-functional theory (SDFT),
which maps the many-body problem in the presence of the external field
${\bf H}_{ext}\r$ on a single-body problem subject to the effective 
field\cite{dftbooks} 
\be
{\bf H}_s\r = {\bf H}_{ext}\r + {\bf H}_{d}\r + {\bf H}_{xc}\r.
\label{hsdef}
\ee
Here the exchange-correlation (xc) magnetic field ${\bf H}_{xc}\r$, 
the SDFT counterpart to the internal exchange field of traditional
theories of magnetism,\cite{herring} is defined as the functional derivative
\be
{\bf H}_{xc}\r = -\frac{\delta E_{xc}[n,{\bf m}]}{\delta \bf m\r},
\label{hxcdef}
\ee
where $E_{xc}[n,{\bf m}]$ is the exchange-correlation functional of
SDFT. ${\bf H}_{d}\r = \nabla \times {\bf A}_d\r$ 
is a Hartree-like term arising from dipolar interactions. 
In the language of SDFT the spin compensation in the JPE would be described 
by saying that, on the average,\cite{footnote1}
\be
{\bf H}_{xc}\r + {\bf H}_{d}\r = - {\bf H}_{ext}\r +\delta {\bf H}\r,
\label{hcancel}
\ee
for some range of densities and temperatures, so that the effective
magnetic field ${\bf H}_s$ vanishes up to at most a small remaining field 
$\delta {\bf H}$ that is not strong enough to destroy superconductivity. 
SDFT does not, however, provide a means to
study the orbital degrees of freedom. Such a means is provided by 
current-density-functional theory (CDFT).\cite{vr1,vr2,vr3}
CDFT is based on the many-body Hamiltonian
\bea
\hat{H}=\hat{T}+\hat{U}+\hat{V} - \int d^3r\, {\bf m}\r \cdot {\bf H}_{ext}\r
\nonumber \\
-\frac{q}{c}\int d^3r\, {\bf j}_p\r \cdot {\bf A}_{ext}\r 
+
\frac{q^2}{2mc^2}\int d^3r\, n\r {\bf A}_{ext}\r^2,
\label{cdftham}
\eea
where $\hat{T}$, $\hat{U}$, and $\hat{V}$ are the operators for the kinetic,
interaction, and potential energy, respectively, $n\r$ is the particle
density, ${\bf m}\r$ the spin magnetization, and ${\bf j}_p\r$ is the
{\it paramagnetic current density}. 
By construction, the CDFT single-particle equations reproduce the current,
particle, and spin densities of the many-body Hamiltonian (\ref{cdftham}),
and this property carries over to the extension of (C)DFT to the
superconducting state.\cite{ogk1,ogk2,asi}
In the CDFT single-particle equations enter three effective potentials: an 
effective magnetic field ${\bf H}_s$, defined as in Eqs.~(\ref{hsdef})
and (\ref{hxcdef})
up to the replacement of $E_{xc}[n,{\bf m}]$ by $E_{xc}[n,{\bf m},{\bf j}_p]$;
an effective scalar potential
\be
V_s\r= v_s\r + \frac{q^2}{2mc^2}
\left({\bf A}_{ext}\r^2 - {\bf A}_s\r^2 \right),
\label{vsdef}
\ee
where $v_s\r =v\r+ v_H\r+v_{xc}\r$ comprises the lattice potential, 
the Hartree potential, and the xc potential;
and finally an effective vector potential, 
\be
{\bf A}_s\r={\bf A}_{ext}\r + {\bf A}_d\r +  {\bf A}_{xc}\r,
\label{asdef}
\ee
whose exchange-correlation contribution is defined as
\be
{\bf A}_{xc}\r = 
-\frac{c}{q}\frac{\delta E_{xc}[n,{\bf m},{\bf j}_p]}{\delta {\bf j}_p\r}.
\label{axcdef}
\ee
${\bf A}_d\r$ describes the dipolar interactions. 
In general these interactions are much weaker than the xc effects,
and one normally neglects ${\bf H}_d\r$ and ${\bf A}_d\r$ in 
Eqs.~(\ref{hsdef}) and (\ref{asdef}).\cite{vr1,vr2,vr3}
However, ${\bf A}_d\r$ includes the Ampere term\cite{vr2} which describes 
the current-current interactions that are responsible for the selfconsistent 
screening of the induced currents in the Meissner phase below 
$H_{c1}$.\cite{ogk2,asi} We thus keep the dipolar terms in the equations. 

Eq.~(\ref{hcancel}) expresses the fact that the basis for the absence 
of Pauli (paramagnetic) pair breaking is
a mutual cancellation between internal and external magnetic fields.
It is tempting to try a similar explanation for the apparent absence of
orbital pair breaking, i.e., assume that 
\be
{\bf A}_{xc}\r +  {\bf A}_d\r  = -{\bf A}_{ext}\r +\delta{\bf A}\r,
\label{acancel}
\ee
where $\delta{\bf A}\r$ is a small term arising from imperfect cancellation,
which is not strong enough to destroy superconductivity. 
Indeed, Eq.~(\ref{acancel}) can be inferred from Eq.~(\ref{hcancel}).
To this end we recall the relation
\be
{\bf H}_{xc}\r = \nabla \times {\bf A}_{xc}\r,
\label{xcrelation}
\ee
which was 
deduced in Ref.~\onlinecite{eschrig} within relativistic DFT and shown in
Ref.~\onlinecite{cdftlett} to hold under very general circumstances 
within CDFT, too. By integrating (\ref{hcancel}) with (\ref{xcrelation}) 
one obtains (\ref{acancel}), up to a physically irrelevant gauge term.
Although (\ref{xcrelation}) can be shown to be an identity under certain
conditions,\cite{eschrig,cdftlett} all one needs for the present
purpose is that it holds to within an accuracy $\delta$, given by the
magnitude of the imperfection in the cancellations (\ref{hcancel})
and (\ref{acancel}). 

The hypothesis put forward in the present paper is, then, that 
Eq.~(\ref{acancel}) is indeed the correct starting point for a
single-particle desription of the JPE, {\it including} the orbital degrees of 
freedom, and that it explains the occurence of a JPE at fields which are close 
to or even above the nominal orbital critical field ${\bf H}_{c2}$.

To further explore this CDFT interpretation of the JPE we provide, in the
remainder of this paper, first a simple illustration of the cancellation 
embodied in Eq.~(\ref{acancel}), then discuss an earlier, closely related, 
theory; next work out some consequences of Eqs.~(\ref{acancel}) and 
(\ref{xcrelation}), and 
finally compare these consequences with experiments and earlier theory on 
the JPE in $\lambda-(BETS)_2FeCl_4$ and $Sn_{1-x}Eu_xMo_6S_8$ 

{\it Illustration of Eq.~(\ref{acancel}): a London superconductor.}
To illustrate in a simple case how Eq.~(\ref{acancel}) explicitly implies
absence of orbital pair breaking, recall the phenomenological London
equation,\cite{tinkham} according to which current and vector potential in
a homogeneous superconductor are related by
$ {\bf j} = - qn_s {\bf A}_{ext}/(mc)$,
where $n_s$ is interpreted as the number density of superconducting electrons.
Since the full physical current is
$ {\bf j} = {\bf j}_p - qn{\bf A}_{ext}/(mc)$,
where the first part is the paramagnetic current, entering the functionals
of CDFT, and the second the diamagnetic current, one obtains for ${\bf j}_p$
\be
{\bf j}_p = \frac{q(n-n_s)}{mc}{\bf A}_{ext}.
\label{jplondon}
\ee
From Eqs.~(\ref{asdef}) and (\ref{acancel}) we obtain 
${\bf A}_s=0 + O(\delta)$, where the terms of order $\delta$ are by assumption 
not strong enough to destroy superconductivity and will be neglected below. 
The CDFT Kohn-Sham equations\cite{vr1,vr2,vr3}
for ${\bf A}_s =0$ necessarily yield ${\bf j}_p=0$. The paramagnetic current
calculated from the CDFT Kohn-Sham equations is, however, by construction
identical with the many-body paramagnetic current.\cite{vr1,vr2,vr3}
When substituted into Eq.~(\ref{jplondon}) (with ${\bf A}_{ext}\neq 0$, since
a nonzero magnetic field is applied), ${\bf j}_p=0$ implies
$n=n_s$, which means that the magnetic field has not broken any Cooper pair
via its coupling to the orbital currents. We thus see explicitly that
Eq.~(\ref{acancel}) implies absence of orbital pair breaking.

{\it Earlier theory:}
A cancellation of orbital effects that is similar (but not identical) to
Eq.~(\ref{acancel}) has been considered already in one of the earliest papers
in the field,\cite{helv1} and rejected as impossible. These investigations were
performed (before the first experimental observation of the JPE) within a
simplified (pre-CDFT) single-particle treatment of the orbital currents.
The authors of Ref.~\onlinecite{helv1} assume that the exchange field
${\bf A}_x$ contained in ${\bf A}_{xc}$ must cancel both,
${\bf A}_d$ and ${\bf A}_{ext}$, and assert that this is impossible since
the exchange interaction is short ranged, while the dipolar interaction
is long ranged. Hence, according to Ref.~\onlinecite{helv1}, no orbital
cancellation can take place. From the present point of view this argument is
not conclusive, because the correlation part of ${\bf A}_{xc}$ may well be 
long ranged since it need not arise from the exchange interaction of
Ref.~\onlinecite{helv1} (explicit examples are given below).
Moreover, all dipolar interactions are suppressed by the relativistic 
prefactor $(v/c)^2$, which makes them much smaller than typical exchange 
effects, so that they can hardly play a decisive role for effects dominated 
by ${\bf A}_{xc}$ and ${\bf H}_{xc}$.
Indeed, in view of the experiments listed above the assumption
that no cancellation of orbital effects takes place seems untenable.

{\it Some consequences of Eq.~(\ref{acancel}):}
First, note that the details of the system do not enter the arguments 
leading to (\ref{acancel}) and (\ref{xcrelation}), which are rather general
and not tied to particular features of the system's electronic structure.
This generality may explain why the JPE could be observed in the physically
very different systems listed above. 

Second, one obtains Eq.~(\ref{acancel}) by integrating Eq.~(\ref{hcancel}). 
Conversely, by taking the curl of Eq.~(\ref{hcancel}) one
recovers, of course, Eq.~(\ref{acancel}). The conditions for spin
compensation and `current compensation' are thus not independent.
\cite{footnote2} In particular, once one has established spin compensation,
current compensation up to some $\delta {\bf A}$ is a consequence. 
In this sense JPE-type\cite{footnote3} 
magnetic-field-induced superconductivity is easier to achieve than is 
commonly thought, because the orbital limitation is less critical than 
it appears on the basis of the original theory.\cite{jp,decrouxfischer} 

Third, the mechanism for spin compensation is located at the
magnetic ions, whose exchange-correlation magnetic field cancels the
external one. Relation (\ref{xcrelation}) implies that the origin for
compensation of the induced currents must be located at the same place.
The search for a physical explanation of current compensation can thus
concentrate on the magnetic ions and, in particular, their correlations.

Fourth, within CDFT ${\bf A}_{xc}$ is not a pure exchange field, but 
encompasses
{\it all} current-related interaction effects. The search for mechanisms by 
means of which current-compensation can be accomplished is thus not 
limited to the conventional exchange interaction. This observation
enables us to consider a wider range of mechanisms (see below), and partially
explains the negative result of Ref.~\onlinecite{helv1} (see discussion of
earlier theory, above).


{\it Experiments on the JPE in $\lambda-(BETS)_2FeCl_4$:}
In connection with very recent experimental work\cite{uji,balicas} on the JPE 
in $\lambda-(BETS)_2FeCl_4$ a simple and explicit physical mechanism was 
proposed\cite{balicas} by means of which the magnetic ions ($Fe^{3+}$) 
inbetween the BETS planes in $\lambda-(BETS)_2FeCl_4$ can simultaneously 
cancel the effects of the external magnetic field on the spin, {\it and} 
suppress the induced orbital currents.
Briefly, Hund's rule correlations imply that the polarized $Fe^{3+}$ ions
have all spin up states of the $Fe$ d-shell occupied, so that these states
are not available as intermediate states for transport of a singlet
Cooper pair from one BETS layer to the next. Currents perpendicular to the 
layers are thus suppressed.\cite{balicas}
Interestingly, this proposal locates the mechanism which suppresses orbital 
pair breaking exactly where it is to be expected on the basis of the above 
arguments, namely at the magnetic ions (cf. consequence three, above). It 
also shows clearly how the magnetic ions can suppress orbital currents by 
interactions different from simple exchange (cf. consequence four, above). 

{\it Experiments on the JPE in $Sn_{1-x}Eu_xMo_6S_8$:}
A different mechanism for the JPE has been proposed in Ref.~\onlinecite{helv2} 
for the ternary chalcogenides, where the JPE was first seen experimentally. 
Here the magnetic $(Eu)$ ions act (apart from producing
spin compensation) as scattering centers, reducing the mean free path and 
thereby increasing $H_{c2}$. This allows the JPE to take place at 
fields above the value of $H_{c2}$ expected in the absence of magnetic 
ions.\cite{fischerSN1,fischerSN2,fischerSN3,helv2} 
Again, we see that the magnetic atoms suppress both, paramagnetic
(Pauli) and diamagnetic (Meissner) pair breaking, and again the
orbital cancellation is not simply due to their exchange interaction with 
the conduction electron orbits.

Two of the three explicit mechanisms proposed in the literature to explain 
the absence of orbital pair breaking in experiments on the JPE are thus 
consistent with the general ideas developed here. To these two one can add
the following alternative scenario for how breakdown of current compensation
can imply breakdown of spin compensation:
As soon as the external magnetic field is strong enough to produce 
spin-polarized currents in the sample, these currents will exert a
torque\cite{spindynlett} on the magnetic ions, which will affect their 
polarization. In extreme cases such torques can lead to a complete 
magnetization reversal,\cite{magrev1,magrev2} but even in less extreme 
situations there is thus a negative feedback between the incipient 
currents and the magnetic ions needed to maintain spin compensation.
Although it is not known at present whether this scenario is realized
in nature, it provides a vivid illustration of the interplay between
orbital effects and spin effects, and of how this interplay can affect
the JPE.

{\it Outlook:}
The main consequence of the above work is that the search for a mechanism
for suppressing orbital currents in JPE-type magnetic-field-induced 
superconductivity is simultaneously narrower and wider than is commonly
thought. Narrower, because one only needs to consider the current-related
effects of the magnetic ions or impurities; wider because these effects are not
limited to simple exchange, but encompass a spectrum of other possibilities.

On a more speculative note, it is worthwhile to point out that the JPE may
well not be the only situation in nature in which similar cancellations take
place, but might constitute a paradigm for other phenomena based on
a complete or partial cancellation between ${\bf A}_{ext}$ and 
${\bf A}_{xc}$. Only further research can show, for example, whether the
transformation of electrons subject to huge magnetic fields into composite
quasiparticles that do not feel any or only a much weaker effective field, 
observed in the fractional quantum Hall effect,\cite{fqhc} can be understood
along similar lines within CDFT. 

{\bf Acknowledgments}
I thank L.~N.~Oliveira, J.~Annett, B.~L.~Gyorffy, and G.~Vignale for useful 
discussions, the FAPESP for financial support, and J.~Quintanilla for bringing 
Ref.~\onlinecite{balicas} to my attention.

\end{document}